\documentclass[a4paper]{article}

\usepackage{INTERSPEECH2022}
\usepackage{xcolor}
\usepackage{booktabs, caption, makecell}
\usepackage{threeparttable}
\title{A Cross-Domain Approach for Continuous Impression Recognition from Dyadic Audio-Visual-Physio Signals}
\name{Yuanchao Li, Catherine Lai}
\address{
  Centre for Speech Technology Research, University of Edinburgh}
\email{y.li-385@sms.ed.ac.uk, c.lai@ed.ac.uk}

\begin{document}

\maketitle
\begin{abstract}
The impression we make on others depends not only on what we say, but also, to a large extent, on how we say it. As a sub-branch of affective computing and social signal processing, impression recognition has proven critical in both human-human conversations and spoken dialogue systems. However, most research has studied impressions only from the signals expressed by the emitter, ignoring the response from the receiver. In this paper, we perform impression recognition using a proposed cross-domain architecture on the dyadic IMPRESSION dataset. This improved architecture makes use of cross-domain attention and regularization. The cross-domain attention consists of intra- and inter-attention mechanisms, which capture intra- and inter-domain relatedness, respectively. The cross-domain regularization includes knowledge distillation and similarity enhancement losses, which strengthen the feature connections between the emitter and receiver. The experimental evaluation verified the effectiveness of our approach. Our approach achieved a concordance correlation coefficient of 0.770 in competence dimension and 0.748 in warmth dimension.
\end{abstract}


\section{Introduction}

Besides the carefully conceived speech content, psychological research indicates that people also intentionally control their appearances and behaviors to leave different impressions on the interaction partners according to different scenarios, such as selection, leader–subordinate relationships, job performance, and so on. For example, a person tends to express extroversion during job interviews and show agreeableness in dating \cite{swider2021first,willis2006first,ambady2008first}. Social research has pointed out the importance of impressions in human interactions, where it is natural to perceive impressions from the partners via non-verbal behaviors, such as eye gaze, body pose, speaking activity, and prosody variation \cite{naumann2009personality,biel2012youtube}. Impressions are often categorized in terms of ``Big Five'' personality traits: extroversion, agreeableness, conscientiousness, neuroticism, and openness. But an impression is more like a formed opinion and some research also uses two dimensions of social perception/cognition: warmth and competence to describe impression. Warmth reflects the intentions of others and includes the meanings such as good-natured, trustworthy, tolerant, friendly, and sincere. Competence reflects the ability of the other to enact his/her intentions and means capable, skillful, intelligent, and confident \cite{cuddy2008warmth,wang2021open}.

The procedure of impression recognition is similar to that of emotion recognition, consisting of two steps: feature extraction and classification/regression. Researchers have used many approaches to predict impressions from human behaviors, such as facial, vocal, and bodily expression. \cite{aran2013one,okada2015personality} used high-level features obtained from body and head motion, speaking activity, along with low-level features extracted from audio to predict the five personality traits. \cite{judge1999big} extracted visual and vocal features (e.g., eye gaze, head pose, speaking activity, and prosody variation) to characterize the social interaction. \cite{subramanian2016ascertain} used Electroencephalogram (EEG), Electrocardiogram (ECG), Galvanic Skin Response (GSR), and facial activity data to recognize personality. Furthermore, human-like speech has been designed for robots and virtual agents, to express different impressions to conduct specific roles. For example, \cite{de2010influence} examined the influence of backchannel selection on extroversion expression using the virtual agent SAL, which acts as an interlocutor in interaction. \cite{yamamoto2018dialogue} analyzed the relationship between personality traits and dialogue behaviors, to control utterance amount, backchannel, filler, and switching pause length for the humanoid robot ERICA to express extroversion, emotional instability, and politeness. However, most research has ignored the fact that impression depends not only on the emitter (system/speaker) but also largely on the perception by the receiver (user/listener). That is, impressions should be formed by both of the interaction partners.

Therefore, in this paper, we propose a dyadic impression recognition approach using cross-domain attention and regularization to capture and strengthen related information from the emitter and receiver as there is a perception gap that separates them into two domains of signal sources (explained in Sec 2.2).

\section{Related Work}
\subsection{Impression Recognition}
The majority of impression research focuses on personality impression. Many corpora have been presented and adopted for personality prediction. YouTube Personality \cite{biel2012youtube} contains 442 vlogs and 2,210 annotations, providing new findings regarding the suitability of collecting personality impressions from crowdsourcing, the types of personality impressions, their association with social attention, and the level of utilization of non-verbal cues. AMIGOS \cite{correa2018amigos} collected participants' EEG, ECG, and GSR signals using wearable sensors, and recorded participants' frontal high definition videos and both RGB and depth full-body videos. Self-assessment of affective levels (valence, arousal, control, familiarity, liking, and basic emotions) were annotated. The SSPNet Speaker Personality \cite{mohammadi2012automatic} consists of 640 speech clips for a total of 322 subjects. Each speech clip lasts 10 seconds and was scored by 11 annotators in terms of the Big Five traits using the BFI-10 questionnaire \cite{gosling2003very,rammstedt2007measuring}.

Compared to personality impression, recognition of warmth and competence is understudied. The Noxi corpus \cite{cafaro2017noxi} was collected to investigate the relationship between observed non-verbal cues and first impression formation of warmth and competence \cite{biancardi2017analyzing}. Noxi is a multilingual dataset of natural dyadic novice-expert interactions, where participants interacted through a screen in different rooms and the experts talked more during the interaction. However, the annotations are from external annotators and do not take into consideration speech content or prosody. Furthermore, there are no physiological signals, eye movements, or self-reported annotations of the receiver.

\subsection{Multimodal Fusion}
Besides corpora, another major issue of impression recognition is the data fusion approach -- a general problem for multimodal affective computing tasks. Traditional early and late modality/feature fusions are being replaced or enhanced by the popular tensor fusion \cite{poria2017review}.
Rather than operations on feature level or decision level, tensor fusion deals with features on the hidden-state level, which can better model synchrony, relatedness, and hierarchy \cite{li2020attention}. For example, attention-based and hierarchical tensor fusion approaches have been investigated for features of different levels and proven useful in emotion recognition \cite{li2021fusing,tian2016recognizing}. However, we regard the fusion problem as more complex in impression recognition for the following reasons: \textit{1) Unlike emotion expression that almost relies on the emitter, impression relies more on the receiver, which means there is a perception gap between the receiver and emitter. 2) Some existing experiments did not conduct real-time interaction, making it difficult for two signal sources to achieve consistency towards the impression.} Thus, we follow prior paths to investigate effective fusion approaches for dyadic impression recognition.

\section{Corpus Description}
\subsection{Dyadic IMPRESSION Dataset}
To address the issues mentioned in Sec 2.1, the dyadic IMPRESSION dataset that contains audio, visual, and physiological signals from both the emitter and receiver has been newly publicized \cite{wang2021open}. The dataset consists of 31 dyads, in total 1,890 minutes of synchronized recordings of face videos, speech clips, eye gaze data, and peripheral nervous system physiological signals. Participants reported their formed impressions in the warmth and competence dimensions in real time. In this work, we use Session 1 of the dataset, where participants (receivers) watched Noxi stimuli (emitters) and annotated the stimuli with respect to warmth and competence. The labels are represented by a stepwise continuous groundtruth, which means the participants were allowed to increase/decrease the label value once they felt an impression change. As a result, there is no value range limitation. Session 1 is separated as a training-validation set and a testing set, which has 40 and 10 participants respectively, and each participant watch 13 stimuli. We ignored the testing set because the labels are not yet provided.

\subsection{Data Processing}
First of all, we selected the features that we need from each modality, as shown in Table~\ref{tab:feats}. Most of the features we selected are the same as \cite{wang2021open}, except for the physiological modality. Some prior research has proved that raw features can have better performance than hand-crafted features \cite{satt2017efficient,li2019improved}. Thus, for physiological signals, we used the raw signals without further feature extraction. We expect the neural networks can extract better representations, considering results reported on extracted hand-crafted features have low correlations with the labels \cite{wang2021open}.

\begin{table}[ht]
\centering
  \caption{Selected features from the emitter and receiver.}
  \label{tab:feats}
  \begin{threeparttable}
  \begin{tabular}{lll}
    \toprule
    Modality & Emitter & Receiver \\
    \midrule
    Audio & MFCC, VoiceProb, & $-$ \\
        & RMS energy, ZCR \\
    Eye & 2D\&3D gaze directions & Gaze duration, \\
        & & 2D gaze locations, \\
        & & 3D eye locations \\
    Facial & AU presence, & AU presence, \\
        & AU intensity & AU intensity \\
    Physio & $-$ & BVP, ECG, GSR \\
    \bottomrule
    \end{tabular}
\begin{tablenotes}\footnotesize
\item[*] AU: Action Units. MFCC: Mel-Frequency Cepstral Coefficients. VoiceProb: Voicing Probability. RMS: Root-Mean-Square. ZCR: Zero-Crossing Rate. BVP: Blood Volume Pulse. ECG: Electrocardiogram. GSR: Galvanic Skin Response.
\end{tablenotes}
\end{threeparttable}
\vspace{-19pt}
\end{table}

Since the sample numbers of each modality are different, we first conducted resampling to make them the same as the label numbers. For the modalities whose sample numbers can be divided by integer, we directly applied the decimate function. Otherwise, the decimate function was first applied to downsample the features to approximate the label sample size. Then a Fast Fourier Transform (FFT) was applied to further downsample the features with a step of \textit{current length/target length}. Since the window is applied every step, it can also serve as a denoising function to remove some outliers. Taking the eye movement modality of the receiver as an example, decimate function with a factor of two or ten (depending on original sample numbers) was directly applied.
For the audio modality of the emitter, decimate function with a factor of three was applied, followed by an FFT with the number of samples that equals the label numbers.
Finally, the 13 emitters have a total of 44,923 samples, each with 412 feature dimensions. The 40 receivers has 44,923 * 40 samples, where each sample has 68 feature dimensions.

\section{Proposed Approach}
\begin{figure*}[ht]
  \centering
  \includegraphics[width=0.82\linewidth]{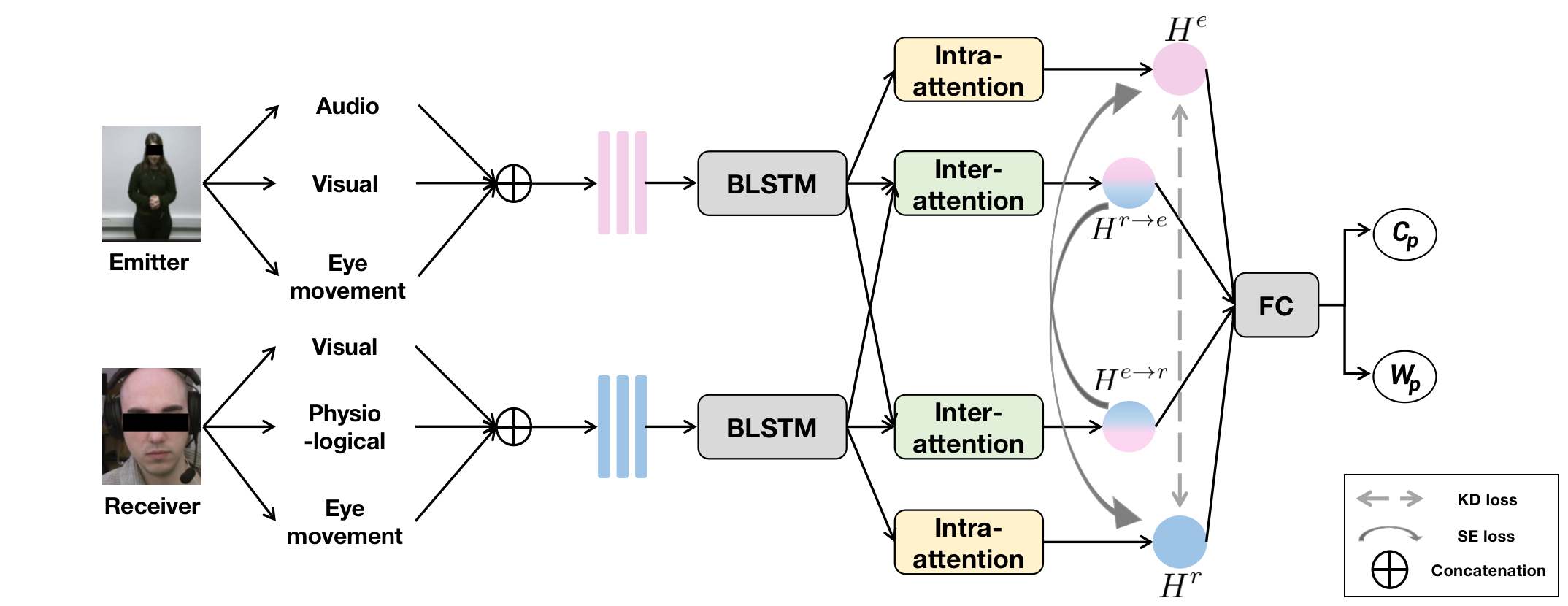}
  \caption{Architecture using the proposed cross-domain approach.}
  \label{fig:model}
\end{figure*}

To address the fusion problem stated in Sec 2.2, we propose a cross-domain approach for dyadic impression recognition, shown in Fig~\ref{fig:model}.

\subsection{Feature Concatenation}
We concatenated the features from the emitter and receiver respectively, using early fusion. Though prior research has proposed State-Of-The-Art (SOTA) tensor fusion approaches for multimodal features \cite{poria2017review}, we used early fusion here because the feature dimension of each modality is relatively small, which is not suitable for use alone in deep learning.

We denoted the concatenated features of the emitter and receiver as $E=\{e_{1}, e_{2}, \cdots, e_{K}\}$ and $R=\{r_{1}, r_{2}, \cdots, r_{L}\}$. $K$ and $L$ are the lengths (dimensions) of the emitter features and the receiver features, which equal to 412 and 68. Then we encoded the inputs $E$ and $R$ as fixed-length vectors $Z^{e}$ and $Z^{r}$.

\subsection{BLSTM Network}
We fed the vectors $Z^{e}$ and $Z^{r}$ to a Bi-directional Long Short-Term Memory (BLSTM) network. The BLSTM network encodes global contexts by updating its hidden states recurrently. It takes as input vectors and outputs a sequence of hidden states $H^{blstm} = \{h^{blstm}_{1}, \cdots, h^{blstm}_{L/n_{pw}}\}$, where $h^{blstm}_{i} \in \mathbb{R}^{2d_{lstm}}$ is the concatenation of the $i$-th forward hidden state and the $i$-th backward hidden state. The forward LSTM reads the input from left to right and the backward LSTM reads the input in reverse.


\subsection{Cross-Domain Attention}
Following the BLSTM network, a structured cross-domain attention consisting of two intra-attention and inter-attention networks, aggregates information from the BLSTM hidden states
$H^{blstm}$, and produce four fixed-length encodings: $H^{e}$, $H^{r}$, $H^{e\rightarrow{r}}$, and $H^{r\rightarrow{e}}$. For both intra- and inter-attention, we used multi-head self-attention mechanism as follows:
\begin{align}
    H &= MultiHead(Q, K, V)W^O \\
        &= Concat(head_1, ..., head_n) 
\end{align}
\vspace{-14pt}
\begin{align}
    head_i &= Attention(QW_i^Q,KW_i^K,VW_i^V) 
\end{align}
\begin{align}
    Attention(Q, K, V) &= Softmax(\frac{QK^{T}} {\sqrt{d}})V
\end{align}
where $W^O$, $W_i^Q$, $W_i^K$, and $W_i^V$ are trainable parameters. $Q$, $K$, and $V$ represent the query, key, and value, respectively, and $d$ is their common dimension. For inter-attention, $Q$ is from one domain (the emitter or receiver), while $K$ and $V$ are from the other. For intra-attention, the three parameters are from the same domain. The value of $n$ is 16, and $H$ denotes the concatenated multi-head representations: $H^{e}$, $H^{r}$, $H^{e\rightarrow{r}}$, and $H^{r\rightarrow{e}}$. Next, we concatenated the four representations and passed them to a Fully-Connected (FC) network with a ReLU activation function to generate the recognition outputs $C_p$ and $W_p$ which are competence and warmth predictions.

We used the cross-domain attention network because impression recognition from the emitter and the receiver can be regarded as different domains in this dataset. In \cite{wang2021open}, the impression labels have high correlations with receiver features, but low correlations with emitter features, indicating that the relatedness between the emitter and receiver is not obvious. This phenomenon is plausible because the receiver responds to the emitter videos without real interaction. Therefore, we need the inter-attention to find relevant features between the two domains.

Inter-attention (also known as co-attention) \cite{lu2019vilbert} has been adopted in affective computing over the recent years \cite{li2021fusing,huang2018speech,huang2021audio}. It exchanges key-value pairs in multi-head self-attention. As shown in Fig~\ref{fig:model}, $H^{e\rightarrow{r}}$ denotes emitter-attended receiver features and $H^{r\rightarrow{e}}$ is the reverse. However, there may be other useful information from individual modalities ignored by the inter-attention. We use intra-attention to resolve this issue. The intra-attention focuses on salient information in each signal domain towards the impression recognition and generates $H^{e}$ and $H^{r}$ as the hidden representations. The four representations are then concatenated for the final non-linear combination.

\subsection{Cross-Domain Regularization}
To reduce the discrepancy and further regulate the relatedness between the two different domains, we propose a cross-domain regularization which has a Knowledge Distillation (KD) loss $\mathcal{L}_{kd}$ and a Similarity Enhancement (SE) loss $\mathcal{L}_{se}$.

\vspace{8pt}

\noindent\textbf{Knowledge Distillation Loss.}
Knowledge distillation is a deep learning technique used for training a small network (student) under the guidance of a trained network (teacher) \cite{romero2014fitnets,tung2019similarity}. Though this technique is widely used in model training, recent work has been inspired to apply it to transfer multimodal knowledge between hidden representations \cite{huang2021audio}. Considering the fact that the multimodal signals from emitter and receiver have weak relatedness, we design a KD loss to transfer the information from the other domain. Unlike the inter-attention which directly attends one domain to the other, which we call ``hard relatedness'', the knowledge distillation enables indirectly learning multimodal knowledge with minor changes in a ``soft'' way: $H^{r}$ and $H^{e}$ can absorb information from each other to some extent while still maintaining their independence. We calculate the Mean Squared Error (MSE) and represent the KD loss as:
\begin{align}
    \mathcal{L}_{kd} &= MSE(H^{e}, H^{r}) + MSE(H^{r}, H^{e}) \\
    MSE &= \frac{1} {m} \sum_{i=1}^m (h^{A}_{i} - h^{B}_{i})^2
\end{align}
where $h^{A}_{i}$, $h^{B}_{i}$ denote the hidden states from two representations, and $m$ is the sequence length.

\vspace{8pt} 

\noindent\textbf{Similarity Enhancement Loss.}
To ensure the inter-attention representations have successfully learned the information from the other domain, we apply a similarity enhancement loss. For example, minimizing the distance between $H^{r\rightarrow{e}}$ and $H^{r}$ to align together the two representations means that the receiver information has been attended to in the emitter information by inter-attention. We use Kullback–Leibler (KL) divergence for this purpose:

\begin{align}
    \mathcal{L}_{se} &= KL(H^{e\rightarrow{r}}, H^{e}) + KL(H^{r\rightarrow{e}}, H^{r}) \\
    KL &= \sum_{i=1}^n P_{A}(H)log\frac{P_{A}(H)} {P_{B}(H)} \\
    P(H) &= \frac{exp(h_{i})} {\sum_{i=1}^m exp(h_{i})}
\end{align}
where $P(H)$ denotes the softened probability vector of the representation $H$, and $m$ is the sequence length.

The reasons that why we choose MSE and KL divergence are: 1) MSE generally outperforms KL divergence in knowledge distillation \cite{kim2021comparing}. 2) KL divergence is good at calculating the distance of two distributions on the same probability space and is popular for similarity measurement \cite{goldberger2003efficient,yao2011symmetric}, so we expect it to enhance the similarity in the cross-domain situation. We also exchanged MSE and KL divergence for KD and SE but found a small decrease in the warmth dimension.

\subsection{Fully-Connected Network}
Finally, the concatenated representations are fed to an FC network containing a linear dense layer with 16 neurons, followed by a ReLU activation function, a dropout layer, and a linear dense layer with a single neuron to generate predictions. The prediction task is optimized by the following objective function:
\begin{align}
    \mathcal{L}_{task} &= MSE(C_{p}, C_{l}) + MSE(W_{p}, W_{l})
\end{align}
where $C_{p}$ and $W_{p}$ are the predictions of competence and warmth, and $C_{l}$ and $W_{l}$ are the corresponding labels.

\section{Experimental Evaluation}
\subsection{Implementation}
The model was built using Pytorch and optimized using the Adam method. The learning rate was set as 1e-3 and reduced by half every 20 epochs. 40 epochs were used for training and validation. The full model was trained by minimizing the overall loss:
\begin{align}
    \mathcal{L} &= \mathcal{L}_{task} + \mathcal{L}_{kd} + \mathcal{L}_{se}
\end{align}
We mixed up all the data and used 80\% for training, 10\% for validation and 10\% for testing. We evaluated the performance using Concordance Correlation Coefficient (CCC).


\subsection{Results and Discussion}
First, we compare our results with the original IMPRESSION paper and present ablation study where we removed each component from concatenation (for the attentions) or from back propagation (for the losses) without changing the architecture. So even when intra-attention was removed, the losses still worked.

\begin{figure}[t]
  \centering
  \includegraphics[width=\linewidth]{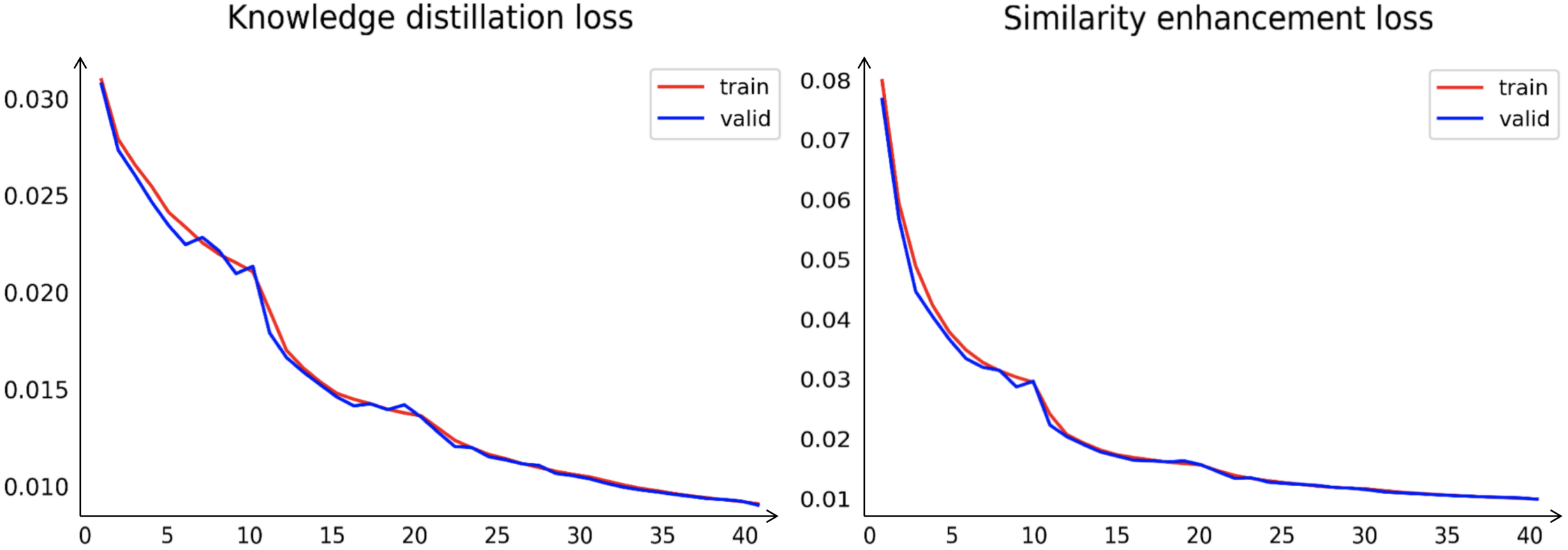}
  \caption{Trends of regularization losses.}
  \label{fig:trend}
\end{figure}

Table~\ref{tab:test} shows that: 1) our proposed full model achieves a similar result in warmth as the original work using LSTM, but a significant increase in competence. 2) The removal of each of the components results in a decrease in performance, which in turn proves their effectiveness. 3) There is no clear difference between removing inter-attention and intra-attention, suggesting that even without the inter-representations, the intra-'s do learn cross-domain relevance with the help of KD loss and SE loss. 4) SE loss shows the least effect among all the components, which is reasonable since the two targets originally have similarities (for instance, $H^{e\rightarrow{r}}$ denotes the receiver information attended by the emitter information which shares similarity with $H^{e}$). On top of that, SE loss reinforces the connection even further.
From Fig~\ref{fig:trend}, we can observe that both KD and SE losses show a decreasing trend. However, they showed an increasing trend (not shown here) when removed from back propagation in the ablation study, which also demonstrates their usefulness.

Next, since the IMPRESSION dataset is just publicized and there are few dimensional impression recognition studies, our results are not directly comparable to the literature. Thus, we compared with some dimensional emotion recognition work. Our prior work \cite{li2017emotion} incorporated sentiment analysis using a weighted linear regression, achieving a CCC of 0.560 in valence estimation. A recent SOTA work \cite{atmaja2020multitask} obtained a CCC of 0.680 for arousal using multi-stage fusion and multi-task learning.


\begin{table}[t]
\centering
  \caption{Ablation study. (-): removal of the component. (E): emitter. (R): receiver.}
  \label{tab:test}
  \begin{tabular}{lcc}
    \toprule
    Model & Competence & Warmth \\
    \midrule
    LSTM \cite{wang2021open} & 0.737 & \textbf{0.751} \\
    Full & \textbf{0.770} & 0.748 \\
    \midrule
    (-) inter-attn. & 0.761 \color{cyan}$\downarrow$0.009 & 0.741 \color{cyan}$\downarrow$0.007\\
    (-) intra-attn. & 0.763 \color{cyan}$\downarrow$0.007 & 0.740 \color{cyan}$\downarrow$0.008\\
    \midrule
    (-) KD loss & 0.763 \color{cyan}$\downarrow$0.007 & 0.743 \color{cyan}$\downarrow$0.005\\
    (-) SE loss & 0.766 \color{cyan}$\downarrow$0.004 & 0.745 \color{cyan}$\downarrow$0.003\\
    \midrule
    (-) facial (E) & 0.729 \color{cyan}$\downarrow$0.041 & 0.717 \color{cyan}$\downarrow$0.031\\
    (-) audio (E) & 0.727 \color{cyan}$\downarrow$0.043 & 0.699 \color{cyan}$\downarrow$0.049\\
    (-) eye (E) & 0.742 \color{cyan}$\downarrow$0.028 & 0.724 \color{cyan}$\downarrow$0.024\\
    \midrule
    (-) facial (R) & 0.678 \color{cyan}$\downarrow$0.092 & 0.644 \color{cyan}$\downarrow$0.104\\
    (-) physio (R) & 0.742 \color{cyan}$\downarrow$0.028 & 0.703 \color{cyan}$\downarrow$0.045\\
    (-) eye (R) & 0.701 \color{cyan}$\downarrow$0.069 & 0.662 \color{cyan}$\downarrow$0.086\\
    \bottomrule
\end{tabular}
\end{table}

\section{Conclusion}
In this paper, we propose a cross-domain architecture to address the dyadic impression recognition problem. This architecture consists of cross-domain attention and regularization to capture and strengthen the relatedness between the emitter and receiver. The cross-domain attention has an intra- and an inter-attention, which focus on the salient information in each domain and the relevant information between two domains. The cross-domain regularization has a knowledge distillation loss and a similarity enhancement loss, ``softly'' transferring multimodal information and minimizing the distance between two domains. The experimental evaluation proves that both components are useful for performance improvement. We expect to test the approach on a second database (e.g., AMIGOS) for more result comparisons in the near future.


\section{Acknowledgements}
The research in this paper uses the IMPRESSION Database collected by SIMS group of University of Geneva, in the scope of IMPRESSION project financially supported by the Swiss National Science Foundation under Grant Number 2000221E-164326 and by ANR IMPRESSSIONS project number ANR-15-CE23-0023.

\bibliographystyle{IEEEtran}

\bibliography{template}

\end{document}